        \documentclass[preprint,12pt]{elsarticle}
  
    \usepackage{titlesec} 

        \usepackage[bottom]{footmisc} 

    \usepackage{graphicx} 
    \usepackage{float} 
    \usepackage{subfigure}
    
        \usepackage{hyperref} 

\begin{document}



\begin{frontmatter}


  \title{Software development practices in academia: a case study comparison%
}

  \author[1,2]{Derek Groen\fnref{fn1}}
  \author[3]{Xiaohu Guo}
  \author[4]{James A. Grogan}
  \author[1]{Ulf D. Schiller}
  \author[5]{James M. Osborne}

  \address[1]{Centre for Computational Science, University College London}
  \address[2]{CoMPLEX, University College London}
  \address[3]{Science and Technology Facilities Council,
Daresbury Laboratory}
  \address[4]{Wolfson Centre for Mathematical Biology, University of Oxford}
  \address[5]{School of Mathematics and Statistics, University of Melbourne}

  \fntext[fn1]{Corresponding author. E-mail: d.groen@ucl.ac.uk, Tel: +44 20 7679 5300}


\begin{abstract}
Academic software development practices often differ from those of commercial development settings, yet only limited research has been conducted on assessing software development practises in academia. Here we present a case study of software development practices in four open-source scientific codes over a period of nine years, characterizing the evolution of their respective development teams, their scientific productivity, and the adoption (or discontinuation) of specific software engineering practises as the team size changes. We show that the transient nature of the development team results in the adoption of different development strategies. We relate measures of publication output to accumulated numbers of developers and find that for the projects considered the time-scale for returns on expended development effort is approximately three years. We discuss the implications of our findings for evaluating the performance of research software development, and in general any computationally oriented scientific project.
\end{abstract}


\end{frontmatter}





\section{Introduction}

Computational simulation and analysis have become integral parts of
scientific research in many disciplines.
Research using large scale simulations is often a collaborative effort
among scientists and engineers, and tends to require the combined use
of multiple software packages developed by different groups. These
efforts often involve a range of algorithmic fields such as
discretization, mesh generation/pre-processing, domain decomposition,
scalable algebraic solvers, statistical analysis, and visualisation.
Given this reliance on software from multiple groups and specialities,
it is important that the user of the software has confidence that the
outcome of the computation is an accurate proxy for the system being
modelled and that it delivers reproducible results. A rigorous quality
assurance system, which helps deliver accuracy and reproducibility, is
a common requirement for software deployment in industry, and it is
increasingly recognised as an essential practice for scientific
software development.  Software engineering approaches form part of a
quality assurance system, and may include methods such as waterfall,
prototyping, iterative and incremental development, spiral
development, rapid application development, and extreme programming
\cite{DeGrace-Stahl:1990,McConnell:1996,Schwaber-Beedle:2001,McConnell:2004}.

Software development practices in academia differ from those in the
commercial sector. In the authors' experience, this is in part due to
the relatively small and transient development teams found in academic
settings, to project-focused development, and to the fact that
academics are incentivized to attract funding and create publications,
and usually care less about financial profit than commercial development. In
addition, academic software development is rarely performed by full
time code developers, and if so, these members are typically hired on
short-term contracts with a focus on a specific subset of the
code. Most scientific software is developed by researchers such as PhD
students and PDRAs (post-docs) who split their time between code development
and research activity. This dichotomy has led to a somewhat reluctant
and heterogeneous adoption of rigorous software engineering practices
in academic contexts.
A survey conducted by Hannay et al.~\cite{Hannay:2009} reveals that
developers of scientific software rely mostly on education from peers
and self-study, and they have relatively little formal training in
software engineering. Moreover, Hannay et al. found that scientists
tend to rank standard software engineering practices higher when they
work on larger software projects.
Heroux et al.~\cite{Heroux:2009} list a number of software practices
which they believe would be easy to adapt and would benefit most
scientific software development teams. They also note that in the
project under consideration, some practices were
only applied to particularly challenging sections of code and relied
on check-lists to make repetitive development and release operations
less error-prone.
Jalali and Wohlin~\cite{Jalali:2010} investigate the adoption of agile
practices in distributed software engineering and find that similar
problems in distributed development have been reported in multiple
articles, possibly pointing to a need to better interpret the context
of different experiences in software development.
Recently, Naguib et al.~\cite{Naguib:2012} note fundamental
differences between software in academia and in industry, i.e.,
academic software is more engineered to achieve high accuracy and
stability, and less so to achieve comprehensibility, maintainability,
and extensibility.
Joppa et al.~\cite{Joppa:2013} have surveyed the attitude towards
scientific software among species distribution modelers, and they
find the ``troubling trend'' that scientists tend to make choices
based on scientifically misplaced factors (e.g., the popularity of the
software-based journal publication, while the software itself has not been
formally been peer-reviewed). They concluded that learning from and
acting on efforts in scientific software development will form an
integral part of delivering models and software that meets high
scientific standards.

In this work, we investigate development practices for four
open-source scientific codes; Chaste\footnote{Chaste:
  \url{http://www.cs.ox.ac.uk/chaste/}}~\cite{PittFrancis:2009,
  Mirams:2013}, HemeLB\footnote{HemeLB:
  \url{https://github.com/UCL/hemelb}}~\cite{Mazzeo:2008,Groen:2013,Groen:2013-3,Nash:2014}
, Fluidity\footnote{Fluidity:
  \url{http://fluidityproject.github.io/}}~\cite{Pain2005,Gorman2004_a,Davies:2011},
and ESPResSo\footnote{ESPResSo:
  \url{http://espressomd.org/}}~\cite{Limbach:2006,Arnold:2013}. These
are codes with which the authors have had development roles, allowing
us to analyse certain aspects of the development process which would
not be accessible otherwise. In our investigation we review the
evolution of the applied practices over the life-time of the codes and
relate them to changes in the development team.  We also discuss
typical practices, such as agile methods with test driven development
and code review, to what extent they have proven effective and have
required modifications over time. Furthermore, we analyse the output
in terms of publications generated using respective software packages
in relation to other factors. In particular, we explore how invested
effort in software development, and adopted practises, translate to
scientific outputs.
To the best of our knowledge, the impact of software development practices on
scientific publication output has not been investigated systematically before,
and our analysis based on the four study cases is the first account in this
direction. Our findings indicate a need to take software development efforts
specifically into account when evaluating the output of science projects that
rely on dedicated software development.

The remainder of this paper is organised as follows; in Section 2 we introduce the
four case studies, along with their specific development practises. In Section 3, we present the methods used to collect the
data on development practices, development team and code size and
publications, and in Section 4 we discuss the changes in development
practices over time. Finally, in Section 5 we investigate the
relationship between code output and expended development effort.

\section{Case Studies}

\subsection{Development, test, and profiling practices}

We briefly summarise the development practices encountered in the four
scientific software projects considered in this work. The purpose here
is not to give a comprehensive overview of software engineering
techniques, but to describe the specific practices that are applied in
the case studies investigated below. The encountered practices also
mirror the prevalent tension between the organizational culture in
academia and the needs of sustainable software development and
maintenance. In the authors' experience, long-term development efforts
are sometimes hampered by short-term funding decisions which typically
do not adequately reflect the potential impact of sustainable
scientific software. The development and testing practices applied in
the studied scientific software projects thus reflect specific ways
chosen by the respective teams to overcome the aforementioned
obstacles.

\subsubsection*{Development tools}

The first step towards sustainable software development typically is
the adoption of tools for version control and release management,
automated build systems, and continuous integration tests. While this
is arguably a rudimentary approach to software engineering, it can
nevertheless lead to considerable improvements in the maintenance of
scientific software.

Version control, or revision control, is used to keep track of all the
modifications made to components of the software project over
time. Revisions can be reviewed, compared, merged and restored
whenever it becomes necessary at a later stage of the project. Version
control systems allow members of the development team to work on the
same files simultaneously, and provide mechanisms to resolve conflicts
that arise from concurrent changes. A variety of modern revision
control systems are available as free software, for example, Mercurial\footnote{\url{https://mercurial.selenic.com}}, SVN\footnote{\url{http://subversion.apache.org/}}, and Git\footnote{\url{https://git-scm.com/}}.

Due to the heterogeneity of modern computing systems and software
environments, compiling source code and linking to the necessary
libraries has become an increasingly involved and time-consuming
task. Build management tools such as the GNU Autotools\footnote{\url{http://www.gnu.org/software/autoconf/}} or CMake\footnote{\url{http://www.cmake.org/}} provide a remedy by offering
automated generation of the necessary steps of the build process, thus
enabling developers to provision their software package on a variety
of different hard- and software platforms.

Continuous integration~\cite{Duvall:2007} involves automated and
periodic execution of unit tests and functional tests, ideally on a
variety of platforms to ensure ease of deployment and correctness of
the software on different platforms.

Scientific software often aims for HPC applications where parallel
performance is of critical importance. Hence performance tuning and
analysis are essential elements of the software development process.
Whereas such aspects are typically not addressed by conventional
software engineering strategies, it becomes increasingly obvious that
scientific codes require performance profiling to be an integral part
of the development and testing process. The Tuning and Analysis
Utilities (TAU)~\cite{Shende:2006} are an example of instrumentation
and measurement techniques that are useful for regular and automated
performance regression tests, which can be referred to as ``continuous
profiling''.

\subsubsection*{Agile methods}

Agile software engineering provides a set of lightweight software
development techniques which are well suited for scientific software
projects due to the flexibility offered and a close connection between
development and usage aspects.

One specific agile methodology referred to in this work is derived
from 'eXtreme programming' \cite{Beck:2004} comprising test-driven
development and pair programming.
Test-driven development involves writing a failing test for a desired
functionality, prior to implementing any source code. Functionality is
then added to the source code until the test passes.  This ensures a
good degree of test coverage for the code.
In this work we distinguish between unit tests and functional tests.
Unit tests cover small units of the code, for example individual
classes and methods, while functional tests are more involved scripts
that use the code as a black box to verify correctness of a certain
functionality. The latter can also be useful to identify sudden
regressions in overall accuracy.
Pair programming~\cite{Cockburn:2000,Williams:2001} is a form of code
review in which two developers simultaneously work on the same code,
and workstation. One developer writes code and the other reviews
it. This helps to capture errors efficiently and also helps less
experienced developers learn the code and development practices.

Another agile methodology that is adopted partially in one of the case
studies is \emph{Scrum} \cite{Schwaber-Beedle:2001}. A key element of
scrum are the basic development units called sprints. Sprints involve
a planned program of work that is then implemented by the developers
during a period of a week up to a month, followed by a review
process. The main objective of a sprint is completion of the work into
a deployable state, including tests and documentation of the
implemented features.

\subsection{Scientific software projects}

\subsubsection*{Chaste}

Chaste (Cancer Heart and Soft Tissue Environment) is a suite of object-oriented
C++ libraries for simulating multiscale processes in biology. Chaste has been
in development, primarily at the University of Oxford, since 2005. 
There are two main applications for Chaste: the first comprises a cardiac
simulation package which provides a way to use electrophysiology models
in tissue simulations using high performance computing. This package has
for example been used to predict a 2-lead human body surface electrocardiogram
(ECG) and its changes following drug administration \citep{Zemzemi:2013}. 
The second application is a multiscale cell based modeling framework which
allows the user to develop simulations of multiple interacting cells using a
variety of agent-based methods including: cellular automata, cell-centre-based
(on and off mesh), cell vertex and cellular Potts models. This framework has
primarily been used for studying cancer in the intestinal crypt, including a
representation of cell mechanics (see e.g.,~\cite{Leeuwen2009Integrative,Osborne2010Hybrid,Dunn2012TwoDimensional}).
For further applications see the overview paper by Mirams et
al.~\cite{Mirams:2013}.

At present, Chaste contains approximately 500,000 lines of C++ code, with seven
regularly (defined in the Methods section) committing developers. The
development team primarily consists of academic researchers (PhD students,
post-docs and research fellows), with dedicated developers occasionally hired
on a per-project basis. 
The development team primarily aim to apply elements of Agile methods
and eXtreme programming. The rationale for the application of these
methods in Chaste is described in Pitt-Francis et
al.~\cite{PittFrancis:2008}, along with a detailed description of each
method.

\subsubsection*{HemeLB}

HemeLB is an open source lattice-Boltzmann simulation environment
which has been largely developed by PhD students and post-doctoral
researchers at University College London. It has been used to model
flow in both cerebral arteries~\cite{Itani:2015} and in retinal
vasculature~\cite{Bernabeu:2014}, and it has six regularly committing
developers as of 2015.
%
The code was first developed in 2005. From 2010 onward, it was
extended with a host of new features and refactored into a more
systematic structure thereby improving its clarity, accuracy and
performance. This effort included the analysis of performance
characteristics in great detail~\cite{Groen:2013,Groen:2013-3} and the
comparison of the accuracy of several boundary conditions and
collision kernels~\cite{Nash:2014}. Current efforts on HemeLB focus on
improving the load balancing of the code, embedding support for deformable
particles, and enhancing the inflow and outflow
conditions~\cite{Itani:2015}.
The software is currently in use across about half a
dozen institutions, primarily in the United Kingdom. 

At present, HemeLB
contains approximately 180,000 lines of code, written largely in heavily
object-oriented and sophisticated C++, though some of the auxiliary tools have
been written in Python.
The development team uses Scrum and employs test-driven development
including unit testing and functional testing within continuous integration.


\subsubsection*{Fluidity}

Fluidity~\cite{Pain2005,Gorman2004_a,piggott2007,Davies:2011} is a
general purpose, multiphase computational fluid dynamics code capable of
numerically solving the Navier-Stokes equation and accompanying field equations
on arbitrary unstructured finite element meshes in one, two and three
dimensions. It is primarily developed at Imperial College London and has been
in development since 1999. 
Fluidity is used in a number of different scientific areas including
geophysical fluid dynamics~\cite{Davies:2011}, computational fluid
dynamics, ocean modeling and mantle convection
(e.g.,~\cite{Pain2005,Hunt:2012}). Fluidity's partial differential
equation simulator employs various finite element and finite volume
discretization methods on unstructured anisotropic adaptive
meshes. The software is parallelized using MPI/OpenMP and is capable
of scaling to tens of thousands of
processors~\cite{Guo2015227,Lange:2013}. Other useful features are a
user-friendly GUI and a Python interface which can be used to validate
the user input, calculate diagnostic fields, set prescribed fields or
user-defined initial and boundary conditions.

At present, Fluidity consists of approximately 500,000 lines of code,
with 12 regularly committing developers.  The developers employ
test-driven development~\cite{pfarrell2011}, and the developement team is currently
integrating TAU~\cite{Shende:2006}
in the agile development process. This serves to obtain performance
feedback for each revision of the code.


\subsubsection*{ESPResSo}

ESPResSo ({\bf E}xtensible {\bf S}imulation {\bf P}ackage for {\bf
  Res}earch on {\bf So}ft Matter~\cite{Limbach:2006,Arnold:2013}) is a
highly versatile software package for performing many-particle
molecular dynamics simulations, with special emphasis on
coarse-grained models as they are used in soft matter research
\cite{Grass:2008,Deserno:2009}. Its development started in 2001 at the
Max Planck Institute for Polymer Research, Mainz, Germany. Since 2010
the ESPResSo project is maintained at the Institute of Computational
Physics, University of Stuttgart, Germany.
ESPResSo is commonly used to simulate systems such as polymers,
colloids, ferro-fluids and biological systems, for example DNA or
lipid membranes. ESPResSo also contains a unique selection of
efficient algorithms for treating Coulomb interactions
\cite{Arnold:2013b,Fahrenberger:2014}. More recently, several grid
based algorithms such as lattice Boltzmann and an electro-kinetic
solver have been implemented as well.
ESPResSo is free, open-source software published under the GNU General Public
License (GPL). It is parallelized using MPI and CUDA and can be employed on
desktop machines, convenience clusters as well as on supercomputers with
hundreds of CPUs/GPUs. The flexibility of the software is enhanced through a
Tcl and Python interface, which allows the user to specify bespoke simulation
protocols.
%

At the time of writing, ESPResSo consists of approximately 220,000
lines of code with 24 regularly committing developers. The
contributors are distributed all over the world and the adoption of
software engineering practices depends largely on individual
commitment. The successful maintenance of ESPResSo relies on the use
of software development tools, e.g., version control, an automated
build system, and continuous integration.

\section{Methods}
\label{Sec:methods}

We collected data for the number of developers and lines of code using
the version control systems of each software. We retrieved this data
from http://www.openhub.net for Chaste, HemeLB and ESPResSo (the
project names there are respectively Chaste, HemeLB and ESPResSo\_MD)
and from legacy version control databases in the case of Fluidity (and
partially for ESPResSo).
We distinguish between three types of developers for each
project. ``Full Time Developers'' are people hired specifically to
develop the software full-time, ``Frequet Research Developers'' are
researchers who have made more than one commit per month, and
``Occasional Research Developers'' are researchers who have made less
than one commit per month, but have made more than one commit in a
given year.
This data was collected from 2005 onwards for Chaste and HemeLB, from 2006
onwards for Fluidity, and from 2001 onwards for ESPResSo.

We assessed the development practices based on our experience working with the
respective codes (JMO and JG with Chaste, DG with HemeLB, XG with Fluidity and
US with ESPResSo) and in consultation with the other members of the respective
development teams. We categorised practices as being: i) regularly or strictly
applied, ii) occasionally or partially applied or iii) rarely or not applied.
We opted for generic descriptions that allow general changes in development
practices to be identified over time and with changes in the size of the
development team, without being overly prescriptive.

The number of publications are based on the number of publications
that could be identified to be directly using the code within a
calender year, with the aim of reflecting a publication list that a
code's website may display. For Chaste and Fluidity a publication list
was available on the code's website. For HemeLB and ESPResSo, the
publications were identified through literature searches on Google
Scholar and Web of Science, followed by manual inspection to filter
out irrelevant results. For ESPResSo, the filtering resulted in
between 50\%-75\% of the Google search results actually being included
in the publication count. The number of publications was collected
from 2005 onwards for Chaste and HemeLB (only data from 2006 onwards
is shown below), from 2006 onwards for ESPResSo and Fluidity.

\section{Results}

\subsection{Development Practices}

\begin{figure}[p]
  \centering
  \subfigure[%
    Chaste.
  ]{%
    \includegraphics[width=0.48\linewidth]{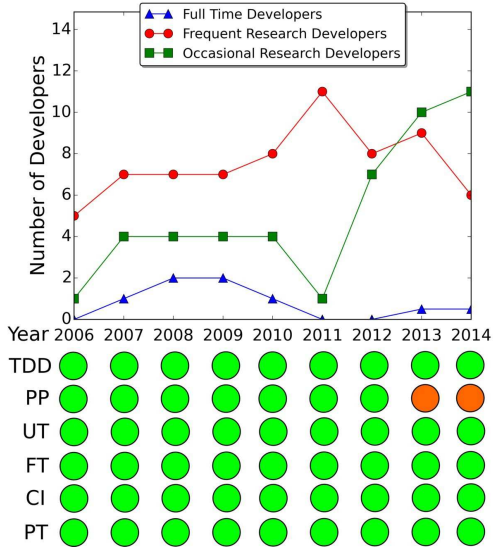}
    \label{Fig:chaste}
  }\hfill%
  \subfigure[%
    HemeLB.
  ]{%
    \includegraphics[width=0.48\linewidth]{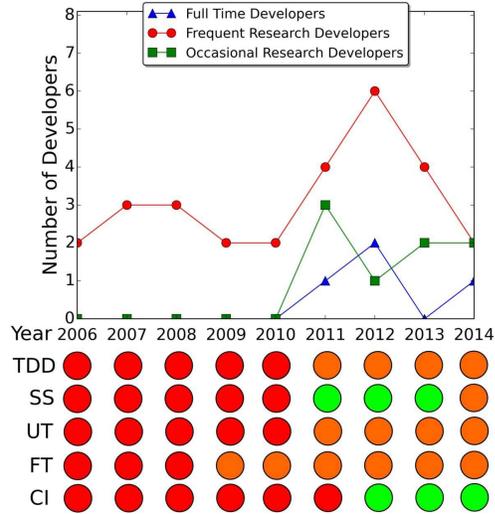}
    \label{Fig:hemelb}
  }\\
  \subfigure[%
    Fluidity.
  ]{%
    \includegraphics[width=0.48\linewidth]{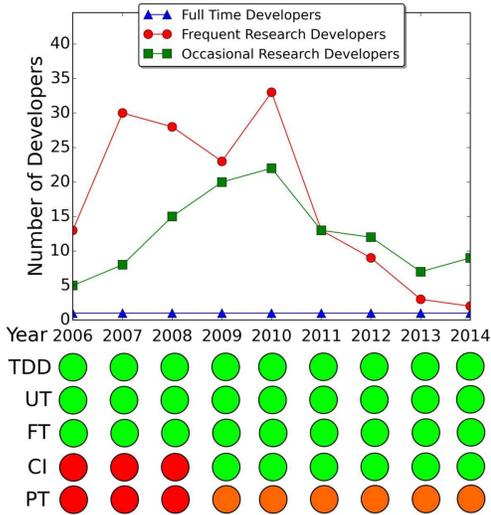}
    \label{Fig:fluidity}
  }\hfill%
  \subfigure[%
    ESPResSo.
  ]{%
    \includegraphics[width=0.48\linewidth]{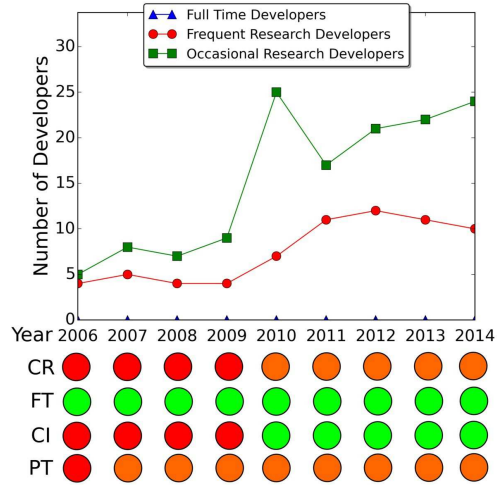}
    \label{Fig:espresso}
  }
  \caption{Changes in development practices with the size and composition of the development teams over the last nine years of development. Key: TDD-Test Driven Development; PP-Pair Programming; UT-Unit Testing; FT- Functional Testing; CI-Continuous Integration; PT-Profiling Testing; SS-Scrum Sessions; and CR-Code Review. Green corresponds to 'always applied', amber is 'partially applied' and red is 'rarely or never applied'.}
  \label{Fig:develcomparison}
\end{figure}

The adopted development practices in relation to the size of the
development teams are shown for the four case studies in
Figure~\ref{Fig:develcomparison}. Green glyphs correspond to regularly or strictly
applied practices, amber to occasionally or partially applied and red
to rarely or not applied practices.

Figure~\ref{Fig:chaste} shows the change in practices for Chaste. The
development team of Chaste strongly emphasises software engineering
practices \cite{PittFrancis:2008}, and aims to apply a consistent set
of engineering practices at all times. The only practice that
sometimes is not applied in full is pair programming. In smaller
development teams, pair programming can become challenging as there
are fewer experts at hand and it can be difficult to find suitable
programmers for a specific pair programming task. However, code
written by a single developer is often reviewed by another developer
before being added to the repository.

Figure~\ref{Fig:hemelb} shows the change in practices for HemeLB. The development team of HemeLB is considerably smaller than those of the other applications and comprises many contributors with a relatively short stay in the team. When several new members joined in 2010 and 2011, the development team was able to adopt a range of well-known engineering practices. Nevertheless, the presence of a large legacy code-base has resulted in various parts not being fully test-covered to this day. In 2014, the development team shrank considerably which resulted in a more loose application of the Scrum system, while many of the testing and development practices are maintained to ensure continued code stability.

Figure~\ref{Fig:fluidity} shows the change in practices for Fluidity. Fluidity is developed and maintained by a large and distributed development team. Besides one constantly hired full time developer since 2006, the Fluidity development team has constantly more than dozens "Frequent Research Developers" until 2011. From 2011 on wards, the "Frequent Research Developers" start turn to "Occasional Research Developers" as the result of Fluidity getting mature. Since 2006, the Fluidity developers start to adapt the automated test-driven development approach, this including systematic unit and functional testing, and continuous integration.

Figure~\ref{Fig:espresso} shows the change in practices for ESPResSo. ESPResSo has a relatively large development team. Besides a core team of three research developers, there is a considerable number of regular contributors and ``remote'' developers. In addition, the research group that maintains ESPResSo relocated twice since the first release. Therefore, it has proven difficult to install consistent software engineering practices, and the extent to which engineering principles are applied depends strongly on personal commitment of the individual developers. Generally, the ESPResSo developers aim to cover every feature by at least one functional test, and since 2010 continuous integration is employed. 

\subsection{Developer Input and Output}

\begin{figure}[p]
  \subfigure[
    Chaste.
  ]{%
    \includegraphics[width=0.49\linewidth]{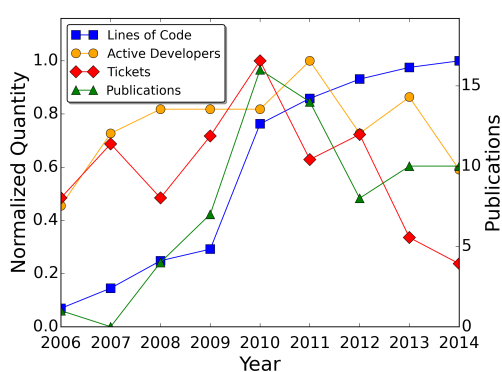}
    \label{Fig:chaste-io}
  }\hfill%
  \subfigure[
    HemeLB.
  ]{%
    \includegraphics[width=0.49\linewidth]{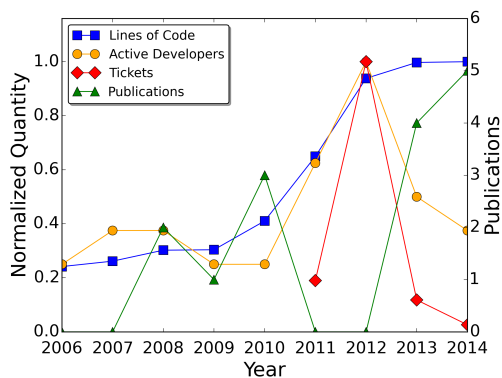}
    \label{Fig:hemelb-io}
  }\\
  \subfigure[
    Fluidity.
  ]{%
    \includegraphics[width=0.49\linewidth]{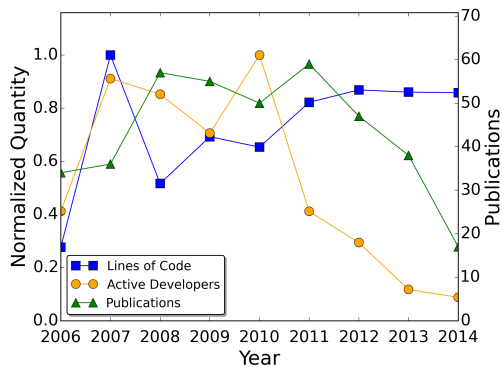}
    \label{Fig:fluidity-io}
  }\hfill%
  \subfigure[
    ESPResSo.
  ]{%
    \includegraphics[width=0.49\linewidth]{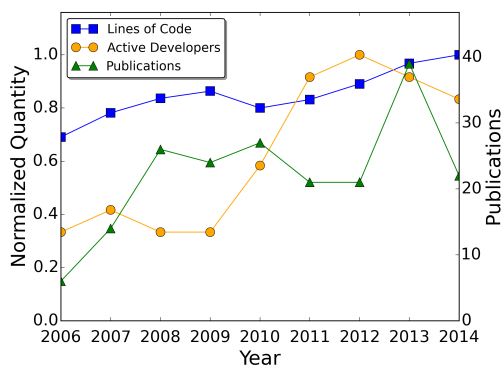}
    \label{Fig:espresso-io}
  }
  \caption{Changes in the size of the development team, code base, number of submitted tickets, and publications over the last nine years of development.}
  \label{Fig:io}
\end{figure}

It is interesting to investigate the publication output of research
software packages in relation to development
effort. Figure~\ref{Fig:io} shows the number of
published papers, code size and the number of active developers
(``Frequent Research Developers'' and ``Full Time Developers'') for
each year from 2006 to 2014. Where applicable, we also provide
information on the number of help tickets raised as a reflection of
code development and use.

For the case of Chaste (Fig.~\ref{Fig:chaste-io}) the number of active
developers has been reasonably constant. The number of
publications per active developer largely increased over the first four years,
decreasing thereafter. The number of tickets raised
shows a behaviour similar to the number of publications.

For the case of HemeLB (Fig.~\ref{Fig:hemelb-io}) the code base size has been closely linked to the size of the development team. In addition, increases in the size of the development team and the introduction of a ticketing system in 2011 are followed by a delayed increase in the number of publications.

For Fluidity in (Fig.~\ref{Fig:fluidity-io}) the code size has been closely related to the active developers and the publications, the code size become more stable when the actively developers and publications goes down.

For ESPResSo (Fig.~\ref{Fig:espresso-io}) the number of active developers has increased considerably between 2009 and 2011, while the number of publications has remained roughly constant. This is in part a consequence of a considerable number of PhD theses and Faraday discussions that were written in 2009 and 2010, prior to the increase in developers. Therefore, the data points in these years lie higher and subsequently screen the increase expected based on the number of active developers.

\subsubsection{Development Effort and Publications}

\begin{figure}[ht!]
\centering\includegraphics[width=0.45\linewidth]{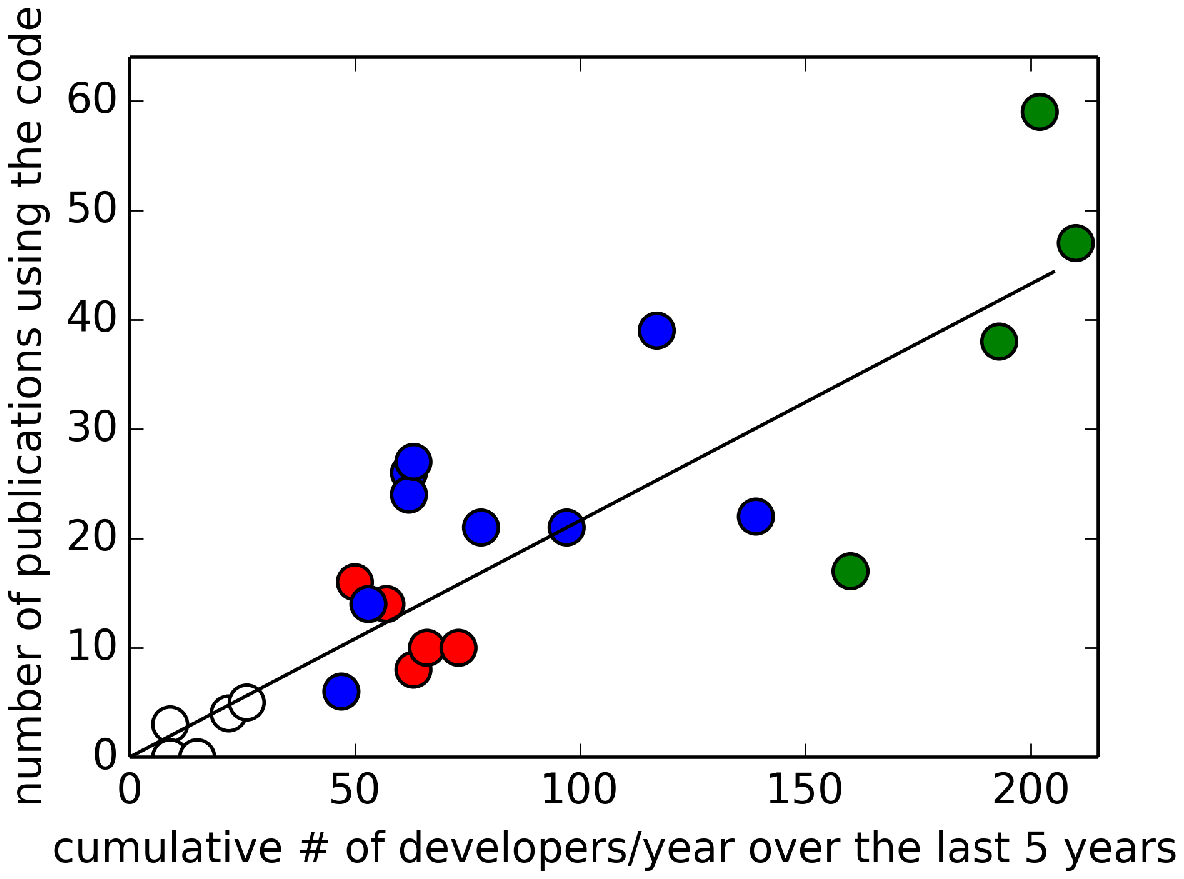}
\includegraphics[width=0.45\linewidth]{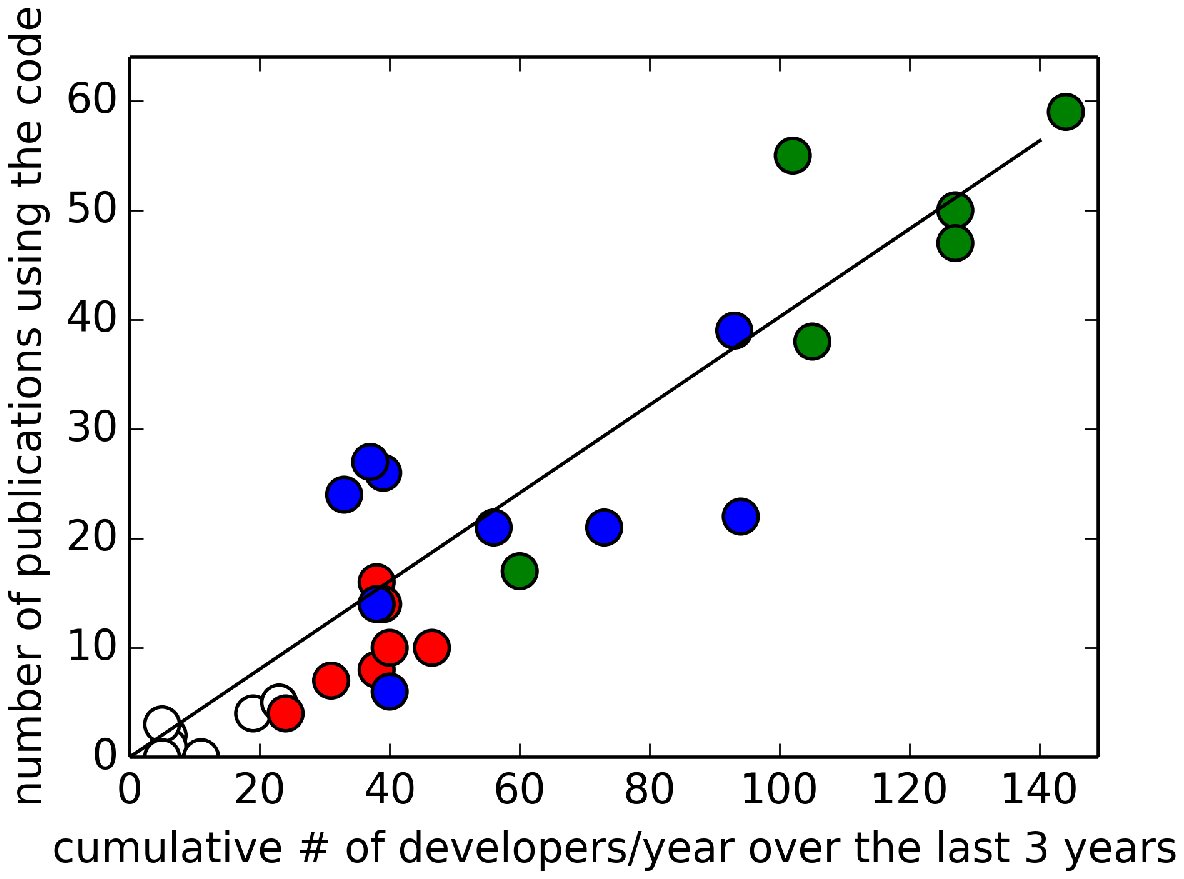}
\caption{Scatter plot of the number of publications in a given year as a function of the cumulative number of developers in the preceding five years (left) and three years (right). Data points are plotted for Chaste (red), HemeLB (in white), Fluidity (green) and Espresso (blue). The black lines are the result of a linear regression on the data sets.}
\label{Fig:pubdev}
\end{figure}

In order to investigate the impact of software development efforts, we
analyze the publication output in relation to the accumulated
development effort. In Fig.~\ref{Fig:pubdev}, we present the number of
publications in a given year as a function of the cumulative number of
developers over the last last five years (a) and three years
(b), respectively, for the four case studies. The scatter plots
show that an increased commitment to software development efforts, as
measured by number of developers, results in more publications based
on these software packages. We performed a linear regression resulting
in a slope of $y = 0.71 \times x$ for the five year data and $y = 0.49
\times x$ for the three year data. Both regressions have $P<0.0001$,
although the regression on the three year data resulted in a higher
R-value ($R=0.86$) than the one on the five year data ($R=0.78$).
Note that in Fig.~\ref{Fig:pubdev} we included the maximum number of
data points available in each case study, which varies between the
software packages due to the different time span for which developer
and publication data was collected (see Sec.~\ref{Sec:methods} for details).

We have also related the number of publications in a given year to the
number of developers in each of the five preceding years,
respectively, and performed a similar regression analysis for all five
relations with results as shown in Tab.~\ref{Tab:pubhist}. We find
that there is a strong effect of the number of developers on the
number of publications which only levels off after three years. In
other words, the software development efforts of our case studies
reveal a lasting impact which can not be appropriately reflected by
measuring publication output on a short term basis. The full merit of
a software project may thus not be accessible until at least three
years after the conclusion of development.

A number of factors have not been taken into account here, such as the
impact of the venues of each publication (e.g., Chaste publications
tend to be less numerous, but quite highly cited) and the exact effort
invested by individual developers (i.e., all contributors are treated
at equal value in this assessment). In order to address the latter, we
have also performed the analysis using a weighting system for
development effort (1.0 for full-time developers, 0.5 for
researcher-developer, and 0.25 for occasional developers) which
resulted in an outcome very similar to the un-weighted assessment
shown here.

\begin{table}[ht!] 
\caption{Relation of historical development effort to publication
  output shown for individual years, ranging from the first year prior
  to publication (first data row) to the fifth year prior to
  publication. For each year we report the slope of the correlation
  found (in publications per active developer that year) in the second
  column, and the obtained P-values, R-values and standard errors
  respectively in the third, fourth and fifth column.}
\label{Tab:pubhist}
\centering 
\begin{tabular}{lllll} 
\hline 
Years ago & slope & P-value & R-value & stderr\\

\hline
1  & 1.11 & 1.7e-11 & 0.879 & 0.11 \\
\hline
2 & 1.07 & 4.2e-10 & 0.884 & 0.10 \\
\hline
3 & 0.99 & 3.8e-8 & 0.824 & 0.13 \\
\hline
4 & 0.71 & 3.0e-4 & 0.652 & 0.17 \\
\hline
5 & 0.73 & 2.1e-3 & 0.607 & 0.21 \\


\end{tabular}
\end{table}

%
%

\section{Discussion}

We conducted and compared four case studies of software development
practices in academia, and analysed the adoption of software
engineering practices, the development effort invested, and the
publication output over a period of at least nine years. We find that
the publication output of the four scientific codes correlates
strongest with the development effort invested in the three years
prior to publication, and that each of these years of effort appear to
contribute equally to the publication output. The correlations become
noticeably weaker when we compare the publication rate with the
development effort invested four or five years earlier.

Based on our results, we conclude that the four considered software
projects should ideally have been reviewed three years after
development efforts have been concluded. This conclusion may be
inconvenient given that performing a final scientific review three
years after the conclusion of a research project can be impractical,
in particular since a large number of academic software development
grants are less than three years in duration. In fact even the
initial efforts of shorter projects cannot be fully assessed by the
final review if the evaluation takes place directly upon the
conclusion of such a project.
There are several ways to mitigate this problem. Firstly, by funding
software projects for at least 3 years reviewers can accurately assess
the scientific publication impact of the initial efforts at the time
when the project finishes. Secondly, review panels could choose to
base their review not directly on peer-reviewed publications, but take
into consideration important preliminary components, such as new
documented code features, simulation data sets and preprints or paper
drafts. Thirdly, for long-running development on academic software,
reviewers could choose to limit their review of the project at hand to
the technical aspects, and judge the academic software on its
scientific potential in a wider context, taking into account the
publication impact of preceding comparable projects.

Although development effort invested in these projects have resulted
in a publication boost, many of the resultant publications do not
feature the original developers as (prominent) authors. For example,
two major contributors to HemeLB only featured on a single
first-author paper each, after investing three years of development
effort. Similarly, a developer contributed the lattice Boltzmann
implementation to ESPResSo which has enabled a host of subsequent
publications of other authors directly using this code. Therefore, the
correlations we presented can not (and must not) be used to assess
individual contributions to the publication output, or to the software
more generally.

Indeed, publication numbers only partially describe the impact of an
individual developer or even a short-term project effort, as they
indirectly reflect more fundamental impact, such as technical quality,
ease-of-reuse and the presence of valuable new features (these metrics
are arguably more difficult to quantify, measure and compare). In
addition, although we find these correlations for our four case
studies, a much larger investigation is required to assess the
relationship between invested development effort and scientific impact
for academic software in general.

In terms of adopted software development practices we find that, in
particular for HemeLB and ESPResSo, new practices are typically
adopted when a development team has recently increased in size. In the
case of HemeLB and Chaste, the application of individual practices was
slightly reduced when the respective development teams became smaller,
but this effect seems to be more limited. More generally, software
development practices, once first applied, appear unlikely to be
abandoned later on. This could indicate a high level of satisfaction
from the development team regarding the adoption of these
practices. 
While a larger study will be required to assess the efficacy of
software engineering practices in computationally oriented research
communities in general, our study indicates that the use of software
engineering principles improves the quality of the research software
and leads to a concomitant increase of the publication output.

\section*{Acknowledgements}

The authors would like to thank: Dr. Tim Greaves, Dr. Matthew Piggott and Dr. Gerard Gorman for help with data collection and interpretation of the Fluidity software package; the Chaste development team for discussions regarding data collection for the Chaste package; and Axel Arnold and Olaf Lenz for valuable
discussions regarding collection and interpretation of data for the
ESPResSo software package. This work was spawned as a collaborative
project during the first 2020 Science \& Software Sustainability
Institute Paper Hackathon which took place in Flore, Northamptonshire,
United Kingdom in September 2014. This work has been funded by the
Computational Horizons in Cancer (CHIC) project under EU FP7 grant
agreement number 600841 (JG), the CRESTA
project under EU FP7 grant agreement number 287703 (DG,US), and the UK
Engineering and Physical Sciences Research Council under grant numbers
EP/I017909/1 (www.2020science.net, DG,JG,JMO) and EP/I034602/1 (US).

\end{document}